\documentstyle[prd,aps,floats]{revtex}

\newcommand{\be}{\begin{equation}}
\newcommand{\ee}{\end{equation}}
\newcommand{\ben}{\begin{eqnarray}}
\newcommand{\een}{\end{eqnarray}}
\newcommand{\bc}{\begin{center}}
\newcommand{\ec}{\end{center}}

\begin{document}

\draft

\input epsf
\renewcommand{\topfraction}{1.0}
\twocolumn[\hsize\textwidth\columnwidth\hsize\csname 
@twocolumnfalse\endcsname

\preprint{SUSSEX-AST 98/?-?, gr-qc/9802258}

\title{\bf A scalar field matter model for dark halos of galaxies
and gravitational redshift}

\author{Franz E.~Schunck}

\address{Institute for Theoretical Physics, University of Cologne,
D-50923 K\"oln, Germany;\\ Astronomy Centre, School of Chemistry, Physics and
Environmental Science, University of Sussex, Falmer, Brighton BN1 9QJ,
United Kingdom}

\date{\today }

\maketitle

\begin{abstract}
We analyze the spherically symmetric Einstein field equation with a
massless complex scalar field. We can use the Newtonian solutions to fit the
rotation curve data of spiral and dwarf galaxies. From the general
relativistic solutions, we can derive high gravitational redshift values.
\end{abstract}


\vskip2pc]

\section {\bf Introduction}

The rotation curves for galaxies or galaxy clusters
should show a Keplerian decrease $v \simeq \sqrt {1/x}$ at the
point where the luminous matter ends. Instead one observes flat
rotation curves beyond the galaxies \cite{peebles}. A linear radial increase
of the mass function of galaxies and galaxy clusters have been derived
from these observations \cite{ost}: $M=v^2_{limit} x$.
Several models have been discussed where
either non-Newtonian gravity \cite{beken} or non-interacting
matter, {\em dark matter} \cite{trimble}, are introduced to solve this problem.
For several classes of gravitational theories, it was recently shown
that the introduction of dark matter is necessary \cite{nester};
cf.~\cite{wein}. Massive compact halo objects, so-called MACHOs,
consisting of baryonic matter are also not able to solve this problem
\cite{alc}.

In 1933 Zwicky \cite{zwi} was the first who suggested the existence
of dark matter in galaxy clusters by investigating the Coma cluster.
The total mass needed to gravitationally bind this cluster exceeds the amount
of the luminous matter by roughly an order of magnitude. Three years later
Smith proved this for the Virgo cluster \cite{smith}.
Beginning of the 1970's one was able to extend the measurements of
the rotation curves of galaxies so that higher mass to luminosity
relations could be found: after some radius the rotation
curves revealed that there is more mass than contained within the
luminous matter \cite{rub}.
The explanation of a linearly increasing mass was first given by Freeman
\cite{fre} providing a spherical halo. The investigations for
determining the radius of a dark matter halo have to go beyond the
HI measurements of the 1980's \cite{sans}, e.g.~by studying satellite
galaxies \cite{zar}, using the weak lensing of background galaxies
by foreground dark halos \cite{brain}, or
looking into quasar absorption lines \cite{webb}; cf.~\cite{via}.
From these investigations, halo radii of more than 200kpc are inferred,
for our Galaxy 230kpc \cite{koc}, and recent results from satellite
galaxies of a set of spiral galaxies show even more than 400kpc
\cite{zar2}. Recently, measurements of rotation curves
of high redshift galaxies have been carried out \cite{vog}.

We present a solution class of a {\em massless complex}
scalar field minimally coupled to the Einstein equation \cite{schunck}.
For the Newtonian types of these solutions, we can fit
rotation curve data of spiral and dwarf galaxies.
The limiting value of the orbital velocity is determined by the central
amplitude of the scalar field. The frequency of the scalar field determines
the halo characteristics of mass and density near the center.

For the general-relativistic (GR) solutions, we can show that they provide
large gravitational redshifts. We discuss how emission and absorption
lines produced in the highly relativistic potential of our model
can be understood. Near the center of
these solutions, we find rotation velocities of about 10$^5$km/s so
that high luminosities can be expected. Spacetime singularities do not appear
within these GR solutions.

A self-gravitationally massive complex scalar field is utilized for the
so-called boson stars \cite{lee,jetzer,kusmartsev,mielke}.
These boson stars have no physical singularities as in our solutions
or in the case of neutron stars. However, real
massless \cite{hehl} or massive \cite{scialom} scalar fields
(in each case without conserved Noether current)
cannot prevent the formation of a singularity (cf.~also the exact solution in
\cite{mann}). Such behavior of singular solutions is supported by
analytical investigations of Christodoulou \cite{chris} and
numerical calculations of Choptuik \cite{chop}.

\section {\bf Einstein-scalar-field equations}

The Lagrange density of a massless complex self-gravitating scalar
field reads
\be
{\cal L} = \frac {1}{2} \sqrt{\mid g \mid} \left [
  \frac {1}{\kappa } R + g^{\mu \nu } (\partial_\mu \Phi^\ast)
  (\partial_\nu \Phi) \right ] \; , \label{lagr}
\ee
where $R$ is the curvature scalar, $\kappa = 8\pi G$, $G$ the
gravitation constant ($\hbar=c=1$),
$g$ the determinant of the metric $g_{\mu \nu }$, and
$\Phi $ the {\em massless complex} scalar field.
Then we find the coupled system
\ben
R_{\mu \nu } - \frac{1}{2} g_{\mu \nu } R  & = &
                  - \kappa T_{\mu \nu } (\Phi ) \; , \\
\Box \Phi & = & 0 \; ,
\een
where
\be
T_{\mu \nu } = (\partial_\mu \Phi^\ast ) (\partial_\nu \Phi )
 - \frac{1}{2} g_{\mu \nu }
 [ g^{\sigma \kappa } (\partial_\sigma \Phi^\ast )
         (\partial_\kappa \Phi ) ]
\ee
is the energy-momentum tensor and
\be
\Box = \partial_\mu
\Bigl [ \sqrt{\mid g \mid } g^{\mu \nu } \partial_\nu \Bigr ]/ 
\sqrt{\mid g \mid }
\ee
the generally covariant d'Alembertian.

For spherically symmetric solutions we use the following static
line element
\be
ds^2 = e^{\nu (r)} dt^2 - e^{\lambda (r)} dr^2
  - r^2 ( d\vartheta^2 + \sin^2\vartheta \, d\varphi^2) \label{metric}
\ee
and, for the scalar field, the ansatz
\be
\Phi (r,t) = P(r) e^{-i \omega t} \; , \\
\ee
where $\omega $ is the frequency of the scalar field.

The non-vanishing components of the energy-mo\-men\-tum tensor are
\ben
T_0{}^0 & = & \rho = - T_1{}^1 = p_r  \nonumber \\
 & = & \frac{1}{2} [ \omega^2  P^2(r) e^{-\nu }
   + P'^2(r) e^{-\lambda } ] \; , \\
T_2{}^2 & = & T_3{}^3 = - p_\bot  \nonumber \\
 & = & - \frac{1}{2} [ \omega^2  P^2(r) e^{-\nu }
   - P'^2(r) e^{-\lambda } ] \; ,
\een
where $'=d/dr$. As equation of state, we find
$\rho = p_r = p_\bot + P'^2(r) e^{-\lambda }$.

The decisive non-vanishing components of the Einstein equation are
\ben
\nu' + \lambda' & = & \kappa (\rho + p_r) r e^\lambda  \; , \label{nula}\\
\lambda' & = & \kappa \rho r e^\lambda - \frac {1}{r} (e^\lambda - 1)
\; , \label{la}
\een
and two further identical components which are fulfilled because of
the Bianchi identities.

The differential equation for the scalar field is
\be
P''(r) + \left ( \frac {\nu' - \lambda'}{2} + \frac {2}{r} \right )
 P'(r) + e^{\lambda - \nu } \omega^2 P(r) = 0  \; . \label{2ska}
\ee

\begin{figure}[t]
\centering 
\leavevmode\epsfysize=7cm \epsfbox{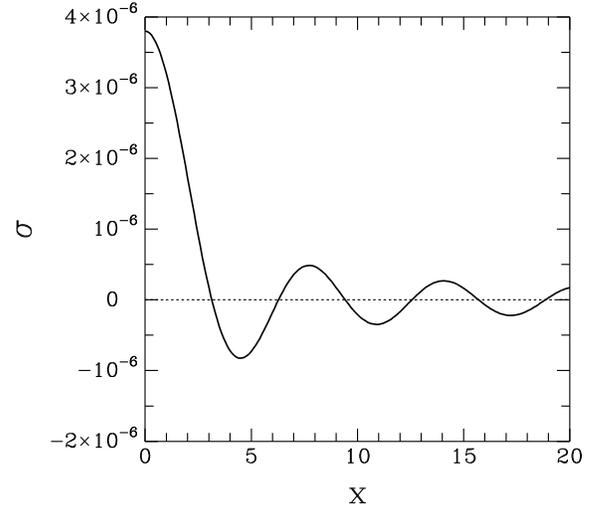}\\ 
\caption[]
{The massless complex scalar field
$\sigma = \sqrt {\kappa /2}\, P$ depending on the dimensionless coordinate
$x=\omega r$ with the initial values
$\sigma (0)=3.8\times 10^{-6}$ $[\sigma (0)(x_{opt})]$ and
$\nu (0)=0$. This solution belongs to the rotation curve with $M_I=-18.29$
in Fig.~\ref{fig3}.}
\label{fig1}
\end{figure}

A typical behavior of the scalar field of Newtonian kind is
demonstrated in Fig.~\ref{fig1}; the metric potentials are almost
constant so that we do not show them here; cf.~Section \ref{newt}.
General relativistic solutions are shown in Section \ref{gene}.

For the rest of our paper, we employ the redefined quantities
$x:=\omega r$ and $\sigma := \sqrt{\kappa /2} \; P$.
The numerical calculation was carried out by
using a Runge-Kutta-routine of the Fortran-libraries IMSL/NAg.
In order to get regular solutions at the origin for the system of
differential equations (\ref{nula}), (\ref{la}), and (\ref{2ska}),
we have to put the initial conditions $\sigma '(0)=0$ and
$\lambda (0)=0$.

The system (\ref{nula}), (\ref{la}), and (\ref{2ska}) possesses the
self-similarity: $x\rightarrow k x$, $\lambda \rightarrow \lambda $,
$e^{\nu }\rightarrow k^{2} e^{\nu }$, and $\sigma \rightarrow \sigma $.
This self-similarity means that
one has only one free parameter, namely the initial value of the
scalar field. One scales the solution simply with the second initial
value for $\nu $.

\section {\bf Newtonian solution}\label{newt}

The Newtonian solutions of our Lagrangian (\ref{lagr}) are characterised by
almost constant metric potentials $\nu ,\lambda $. The scalar field equation
(\ref{2ska}) can then be rewritten into the following form:
\be
\sigma '' + \frac{2 \sigma '}{x} + \sigma = 0 \; , \\
\ee
($'=d/dx$) which has the solution
\be
\sigma (x) = \frac{1}{x} \left [ A \sin (x) + B \cos (x) \right ] \; ,
\label{sigma}
\ee
where $A,B$ are some constants. For $B\neq 0$, the solution is singular at the
origin, why we can rule out it.

The Newtonian form of the energy density reads
\be
\rho (x) =  \frac{A^2}{x^2} \left [ 1 - \frac{\sin (2 x)}{x}
+ \frac{\sin^2 (x)}{x^2} \right ]  \label{rho} \; .
\ee
For small $x$, we have $\rho \propto A^2 [1 - 2 x^2/9 + x^4/45]$,
i.e.~it is proportional to a constant. For dwarf galaxies,
one observes such a behavior of the density \cite{moore} where
good fitting results are obtained by using the empirical
isothermal density profile
$\rho (x) \propto 1/(x_c^2+x^2)$, where $x_c$ is the core radius.
Hence, our constant $A$ resembles a core radius.
By applying a maximum halo model, i.e.~without
baryonic matter, the value of $A$ is higher than for a model
including baryonic matter. Therefore the core radius increases if one
adds baryonic matter as found in \cite{moore}; cf.~\cite{flo}.

\begin{figure}[t]
\centering 
\leavevmode\epsfysize=7cm \epsfbox{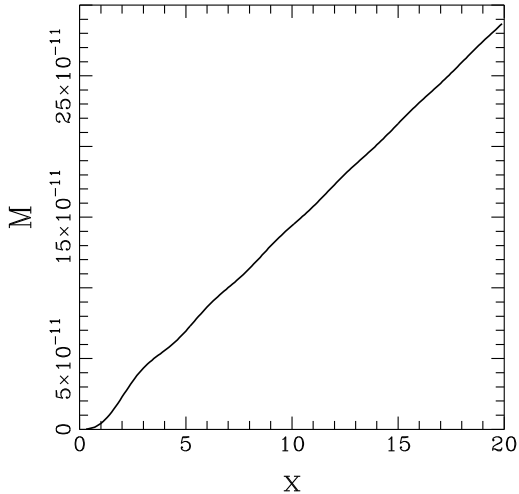}\\ 
\caption[]
{The mass function in units of $[\sigma^2 (0)(x_{opt})/(\omega \kappa )]$
increases near the center like $x^3$ and at higher radial values linearly.
We use here the same initial values as in Fig.~\ref{fig1}, so that this curve
shows the behavior of mass for the rotation curve with $M_I=-18.29$
in Fig.~\ref{fig3}.}
\label{fig2}
\end{figure}

The general solution of Eq.~(\ref{la}) is
\be
e^{-\lambda } = 1 - \frac{M(x)}{x}  \label{lamb}
\ee
with the mass function
$M(x) = \int_0^x \rho (\zeta ) \zeta^2 d\zeta $.
We find the Newtonian formula (cf.~Fig.~\ref{fig2})
\be
M(x) =  A^2 \left \{ x + \frac{\cos (2 x) - 1}{2 x} \right \} 
 \label{mass} \; .
\ee
Comparing with the expected Newtonian result $M=v^2_{limit}x$, we see
that $v_{limit}=A$; hence, the amplitude of the scalar field at the
origin determines the limiting orbital velocity.
For small $x$, the mass function behaves like
\be
M(x) \propto A^2 x^3/3 - 2A^2 x^5/45 + {\cal O}(x^7) \; .
\ee
This corresponds
to the constant density at the center. At higher radial
distances, the mass function shows the linear behavior.

Asymptotically, following (\ref{lamb}), the metric potential
$e^{- \lambda }$ approaches the value
$C^2 := 1 - A^2$, where $C^2<1$. After a redefinition of the coordinate
$x \rightarrow C^{-1} x \; ,$
the asymptotic space has a deficit solid angle. The area of a sphere of
radius $x$ is not $4 \pi x^2$, but $4 \pi C^2 x^2$; cf., e.g., analogous
results for global monopoles and global textures \cite{barr} where one
also finds a linear increase of the mass function.
We show in Sections \ref{cons} and \ref{cover} how one can find a
closure of the solutions and avoids this asymptotic problem.

Following (\ref{nula}), the second metric potential behaves
asymptotically like $e^\nu  \rightarrow  x^K$,
where $K = 2 A^2/C^2 = 2 (1/C^2 - 1) >0$. The behavior of both metric
potentials could be confirmed numerically.

A question is the validity of the formulas (\ref{sigma}), (\ref{rho}),
and (\ref{mass}). We found for $\sigma (0)=10^{-3}$ and $\nu(0)=0$
at $x=10$ a deviation of 0.001\%, at $x=100$
of 0.3\% and at $x=1000$ of about 4\% . This shows clearly that these
formulas should be used only near the center. All figures in this paper
were produced by using the numerical solutions. If one is interested
into the asymptotic behavior of the solutions, the formulas can still
be used {\em asymptotically} (e.g.~for the calculation in the Tables)
because one can still derive the order of magnitude from them as we confirmed
numerically. For higher intial values of $\sigma $, only numerically
determined solutions can be used.

\section {\bf Rotation curves}

In this Section we shall model rotation curves of dwarf and spiral
galaxies. Observations show that rotation curves are becoming flat
(constant orbital velocity) in the surrounding region of galaxies where data
are received from the 21cm wavelength of neutral hydrogen (HI).
But the mass density of the neutral hydrogen is not sufficient to explain
this velocity behavior. Therefore, we introduce dark matter consisting of a
massless complex scalar field which
interacts with the luminous matter exclusively by the gravitational force.
We apply only Newtonian solutions of our model.

\begin{figure}[ht]
\centering 
\leavevmode\epsfysize=5cm \epsfbox{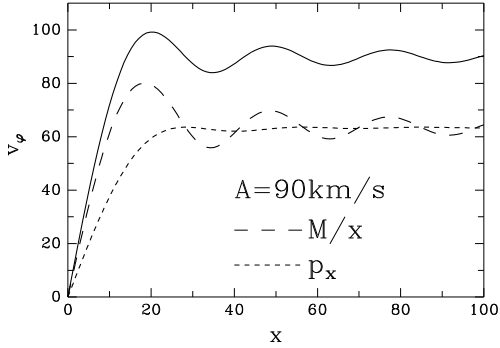}\\ 
\caption[]
{A generic rotation curve for an initial value of the scalar field
$\sigma (0)=A=3\times 10^{-3}$. The Newtonian (M/x) and the pressure
(p$_x$) contributions are shown separately. In the rotation curve fits,
mainly the first part up to the maximum is used. The rotation curve
oscillates around the asymptotic value of 90km/s.
The velocity $v$ is measured in [km/s] while the dimension of the
radial coordinate $r$ depends on the choice of $\omega $; $x$ is
dimensionless.}
\label{fig3a}
\end{figure}

For the static spherically symmetric metric (\ref{metric}) considered
here, circular orbit geodesics obey
\ben
v_\varphi^2 = \frac{1}{2} r \nu' e^{\nu } & = &
\frac{1}{2} e^{\nu } (e^{\lambda }-1) +
\frac{1}{2} \kappa p_r r^2 e^{\lambda +\nu } \nonumber \\
 & \simeq &
\frac{M(r)}{r} + \frac{1}{2} \kappa p_r r^2 e^{\lambda +\nu }
\label{vphi}
\een
which reduces outside of matter for a weakly gravitational field into the
Newtonian form $v_{\varphi, Newt}^2 = M(r)/r$. But, for $p_r \neq 0$,
we have to use the general-relativistic formula.
By using the Newtonian solutions of Section \ref{newt}
within (\ref{vphi}), we find
\be
v_\varphi^2 = A^2 \left [ 1 - \frac{\sin(2x)}{2x} \right ] \; ;
\ee
cf.~a generic rotation curve in Fig.~\ref{fig3a}.

Asymptotically, from (\ref{vphi}), we have
\be
v_\varphi^2=e^{\nu} [ A^2/2 + A^2/2 ]
\ee
what means that both the first Newtonian part $e^{\nu } (e^{\lambda }-1)/2$
and the second matter part contribute the same amount to the
rotation velocity.
For $e^{\nu}=1$, we have $v_\varphi=A$ as we had already derived from the
general Newtonian solution for the mass function.

The next step is now to take observational data for spiral and dwarf
galaxies and try to model them by using our model and a model which describes
the luminous matter distribution.
For spiral galaxies, we use both the universal rotation curves of
Persic, Salucci, and Stel \cite{per} and some individual ones.
Persic et al.~confirmed by
investigating data of 967 spirals that the structural properties of
dark and visible matter are linked together. This means that a
spiral galaxy with low luminosity is stronger dominated by a dark matter
halo than a spiral one with high luminosity.
This is also one possible statement
of the Tully-Fisher relation \cite{peebles}. From this, it follows
that a low luminosity spiral has a rather increasing rotation curve
and a high luminosity spiral galaxy a rather decreasing rotation curve.

\begin{figure}[ht]
\centering 
\leavevmode\epsfysize=12cm \epsfbox{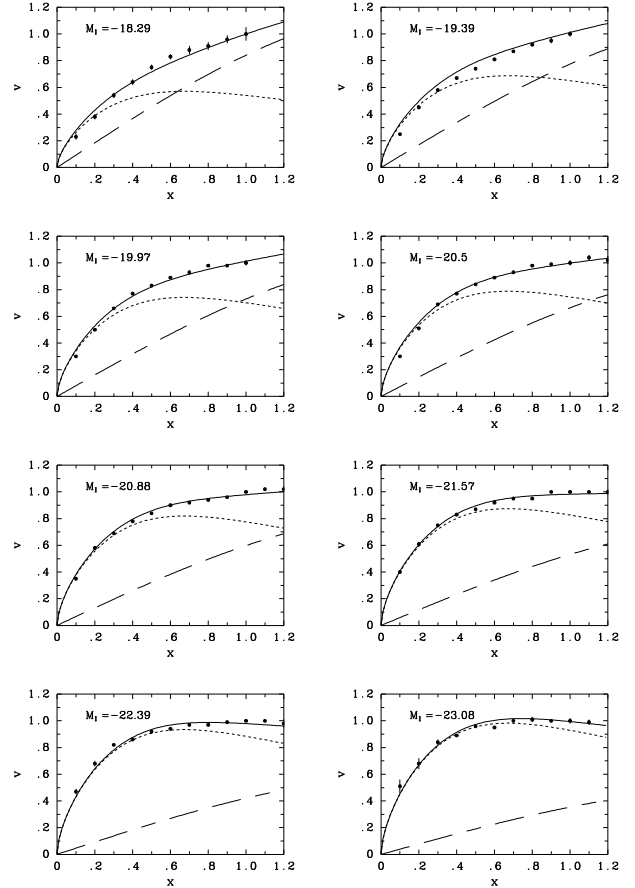}\\ 
\caption[]
{Eight universal rotation curves with different absolute magnitudes
$M_I$ introduced in \cite{per}. The radial coordinate $x$ in units of the
optical radius $x_{opt}$ which encompasses 83\% of the total integrated
light and the velocity $v$ in units of $v(x_{opt})$. For the halo (long-dashed
curve), the initial values are:
$\sigma (0)={3.8,3.5,3.3,3.1,2.7,2.4,1.9,1.6}\times 10^{-6}$ of $M_I=-18.29$
for the first and $M_I=-23.08$ for the last value.}
\label{fig3}
\end{figure}

In the following, we model the universal rotation curves by a combination
of a stellar disk and our halo. The rotation curve for the stellar disk
follows from an exponential thin disk light distribution \cite{per}
\ben
& & v_{disk}^2(x)/v^2(x_{opt}) \nonumber \\
& & \; = \left [ 0.72 + \frac {0.325}{2.5} (M_B^\ast - M_B)
  \right ] \frac{1.97 x^{1.22}}{(x^2 + 0.78^2)^{1.43}} \; , \label{pers}
\een
where $M_B^\ast = -20.5$ is the absolute magnitude in the blue band
(corresponding to $\log L_\ast =10.4$) and $M_B = - 0.38 + 0.92 M_I$
($M_I$ from the $I$ band); this formula can be used within the range
$0.04 \simeq x/x_{opt} \le 2$. The $x$ depending factor arrives from
approximations of modified Bessel functions (see below); the constants in the
$M_B$ depending factor are arranged that it gives the best fit
for our universal rotation curves; cf.~\cite{per}.
The total rotation curves result then from
\be
v_{total} = \sqrt {v_{disk}^2 + v_{halo}^2} \; .
\label{total}
\ee
The outcome can be seen in Fig.~\ref{fig3}. One recognizes a very good
agreement of the data.

From this fit, we find that the amount of disk matter has to decrease with
decreasing luminosity, i.e.~decreasing absolute magnitude.
This is a consequence of the Tully-Fisher relation,
but is verified here by fitting the data. From \cite{per}, it follows
that the rotation curves are self-similar, i.e.~only one parameter,
the luminosity or the limiting rotation velocity,
establishes completely the properties of the halo far from the luminous matter
region. This property is also revealed in our model where asymptotically
only the parameter $A$ determines the halo; cf.~the Newtonian solutions of
Section \ref{newt}. Moreover, we can connect this
one free parameter with the amplitude of the scalar field at the center.

\begin{figure}[t]
\centering 
\leavevmode\epsfysize=8.5cm \epsfbox{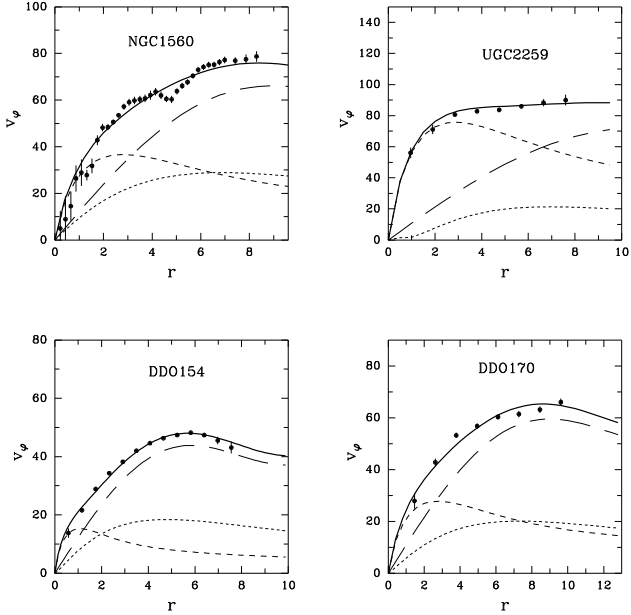}\\ 
\caption[]
{Dwarf galaxies: Rotation curve fits for NGC1560, UGC2259, DDO154, and
DDO170 with halo (long-dashed), stars (dashed), and HI gas (short-dashed).
The velocity $v$ is measured in [km/s] and the radial coordinate $r$ in
[kpc]. The parameters for these fits are summarized in Table \ref{table7}.}
\label{fig4}
\end{figure}

With the improvements of measuring rotation curves, HI rotation curve data
have been found, investigating higher distances from the galaxy center.
For an exponentially thin matter disk
with a surface density $\Sigma (x/x_0)=\Sigma_0 \exp (-x/x_0)$, where
$x_0$ is the disk scale length and $\Sigma_0$ some constant at $x_0$,
the contribution to the rotation velocity can be determined by \cite{fre}
(cf.~also \cite{car2})
\ben
& & v_{disk,gas}^2 (r) = \frac{\kappa \Sigma_0 r^2}{8r_0} \times \nonumber \\
& &
\left [ I_0 \left (\frac{r}{2r_0} \right ) K_0 \left (\frac{r}{2r_0} \right )
- I_1 \left (\frac{r}{2r_0} \right ) K_1 \left (\frac{r}{2r_0} \right )
\right ] \, ,
\een
where $I_n$ and $K_n$ are modified Bessel functions. Equation (\ref{pers})
is actually the approximation of this exact result. This formula shall be
used in the following for the contribution of stars and gas in individual
galaxies, with specific constants $r_0$ and $\Sigma_0$. The gas part
contributes with a summand $v_{gas}^2$ under the square root in
(\ref{total}).

In some galaxies, a clear bulge can be read off the light curve.
For these cases, we use the formula from Kent \cite{kent} in order to
calculate the circular velocities from the observed surface density
$\sigma_{bulge}(r)$
\ben
& & v_{bulge}^2 (r) = \frac{\kappa}{4r} \int\limits_0^r
\zeta \sigma_{bulge}(\zeta ) d\zeta  + \nonumber \\
& & \frac{\kappa}{2\pi r} \int\limits_r^\infty \left [
\arcsin \left (\frac{r}{\zeta } \right )- \frac{r}{\sqrt{\zeta^2-r^2}} \right ]
\zeta \sigma_{bulge}(\zeta ) d\zeta  \, .
\een
Corrections for flattening of bulges are ignored by this formula.

The galaxies we choose belong to a sample of 11 galaxies of different Hubble
types and absolute magnitudes which fulfill several strong requirements
\cite{beg}. All rotation curve data are measured in
the 21-cm line of neutral hydrogen, so that the gas distribution
extends far beyond the optical disc (at least 8 scale lengths)
and the necessity of dark matter is becoming obvious. Isolation of
galaxies are another constraint so that perturbative effects of
nearby situated galaxies are negligible. Besides of that, high quality
data are required.

\begin{figure}[t]
\centering 
\leavevmode\epsfysize=7cm \epsfbox{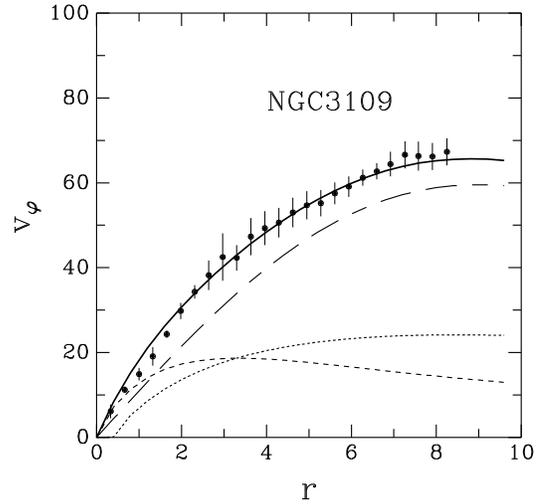}\\ 
\caption[]
{Dwarf galaxy: Rotation curve fit for NGC3109
with halo (long-dashed), stars (dashed), and HI gas (short-dashed).
The velocity $v$ is measured in [km/s] and the radial coordinate $r$ in
[kpc]. The parameters for this fit are summarized in Table \ref{table7}.}
\label{fig5}
\end{figure}

For the calculation of the neutral hydrogen gas for the galaxies
of this sample, a further remark is in
order. In \cite{man2}, it was shown that the HI data can be decomposed
into a sum of two or three exponential disks. We shall use the results
of this paper. We start now fitting dwarf galaxies.

\begin{table}[bth]
\caption[]{Data for rotation curve fits: luminosity L, disk scale length
$r_0$, scalar field frequency $\omega $, mass-to-light ratio for disk and
bulge, initial value of the scalar field $\sigma $. In all cases, we have used
$\nu (0)=-5\times 10^{-6}$. The parameters for the decomposed HI gas
surface densities can be found in \cite{man2}; we have multiplied the
total NGC3109 HI mass by a factor of 1.67 because of some missing 21 cm
line flux at the VLA where the NGC3109 observations were made \cite{job,beg}.
[Hubble parameter $H_0=75$ km/(s Mpc)].}
\begin{tabular}{d|ccccccc}
  & L & \ \ $r_0$ \ \ & \ \ $\omega^{-1}$ \ \ &
(M/L)$_{\rm d}$ &
(M/L)$_{\rm b}$ & \ \ $\sigma (0)$ \ \ \\
  & [10$^9$M$_\odot$] & [kpc] & [kpc] & & & [$10^{-4}$] \\
\tableline
DDO154  & 0.05 & 0.50 & 2.6  & 1.4  & ---  & 1.325 \\
DDO170  & 0.16 & 1.28 & 4.0  & 3.7  & ---  & 1.8   \\
NGC1560 & 0.35 & 1.30 & 4.0  & 3.0  & ---  & 2.0   \\
UGC2259 & 1.02 & 1.33 & 5.0  & 4.5  & ---  & 2.2   \\
NGC2403 & 7.90 & 2.05 & 9.0  & 2.0  & ---  & 3.5   \\
NGC2903 & 15.3 & 2.02 & 9.0  & 3.5  & ---  & 4.4   \\
NGC2998 & 90.0 & 5.40 & 1.8  & 0.25 & 1.5  & 5.9   \\
NGC3109 & 0.81 & 1.55 & 4.0  & 0.4  & ---  & 1.8   \\
NGC3198 & 9.00 & 2.63 & 14.0 & 3.83 & ---  & 3.87  \\
NGC7331 & 54.0 & 4.48 & 20.0 & 2.25 & 1.7  & 6.0   \\
\end{tabular}
\label{table7}
\end{table}

Dwarf galaxies have the characteristic of very low luminosities \cite{moore}.
Following empirical models, their rotation curves vary
from the ones of spiral galaxies in such a way
that instead of the Hernquist profile \cite{hern} an isothermal
density profile has to be applied \cite{moore}; cf.~also \cite{puc}.
From the rotation curves, it can be derived that the dark matter density
near the center is almost constant, i.e.~dark matter has a core in
these galaxies. Our Newtonian solutions
for the massless complex scalar field reveal a constant density
near the center, so it is not surprisingly that
good fits for five dwarf galaxies from the Begeman et al.~sample
have been found (Fig.~\ref{fig4} and \ref{fig5}) \cite{lak,bro,job,bege}.
In all cases, it is recognizable that the dark matter halo dominates
the luminous parts of the galaxies. Especially
remarkable is the maximum of the DDO154 data which can be matched
perfectly by our dark matter halo; hence, decreasing
rotation curve data, at the end of observational resolution,
can be explained by using only a dominating dark matter component.
The prediction of our model is to find oscillations in rotation curve
data at high distances from the center of galaxies.

\begin{figure}[t]
\centering 
\leavevmode\epsfysize=8.5cm \epsfbox{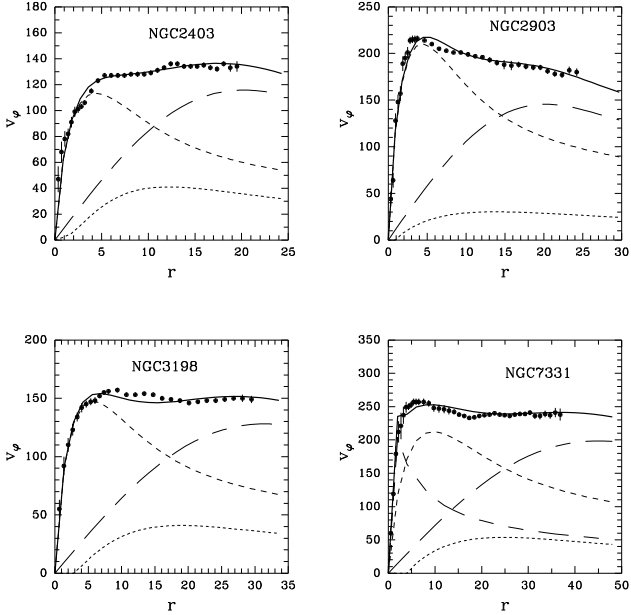}\\ 
\caption[]
{Spiral galaxies: Rotation curve fits for NGC2403, NGC2903, NGC3198,
and NGC7331 with halo (long-dashed), stars (dashed), and HI gas (short-dashed);
NGC7331 has also a bulge included which is truncated at 6.2kpc \cite{bege}.
The velocity $v$ is measured in [km/s] and the radial coordinate $r$ in
[kpc]. The parameters for this fit are summarized in Table \ref{table7}.}
\label{fig4a}
\end{figure}

Our results for the sample of spiral galaxies are summarized in
Fig.~\ref{fig4a}, cf.~\cite{bege,van}. In all cases, we are able
to produce good fits. Models using an approximate isothermal sphere or
a modified Newtonian dynamics (MOND) were also applied for these
galaxies, e.g.~\cite{ken,beg}. Conformal gravitation theory
has been explored in \cite{mann,man1,man2,man3}; the
disadvantage of that model is its increasing rotation curves.

\begin{figure}[t]
\centering 
\leavevmode\epsfysize=7cm \epsfbox{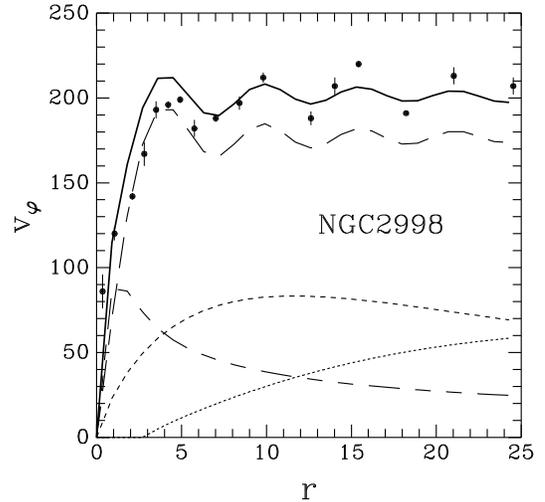}\\ 
\caption[]
{Spiral galaxy with oscillating rotation curves: NGC2998
with dominating halo (long-dashed), stars (short-dashed), HI gas (dotted),
and bulge (dashed). The velocity $v$ is given
in [km/s] and the radial coordinate $r$ in units of [kpc].
In order to find a good correspondence with the data
it was necessary to have a dominating halo.
The extrema of data and halo are at about the same values of $r$.
The bulge is truncated at 2.9kpc; the data for the bulge are taken
from \cite{kent}, but notice a different value for the Hubble constant there.
The HI gas contribution \cite{broeils}
was decomposed into $\sigma_{HI}=55 \exp(-r/12\mbox{kpc})-
49 \exp(-r/9\mbox{kpc})$ [M$_\odot$/pc$^2$].
For other parameters, see Table \ref{table7}.}
\label{fig5a}
\end{figure}

By building universal rotation curves, a smoothing method is
used \cite{per}. Important informations about the distribution of
the dark matter within individual rotation curves could be lost by this
method. One can find an oscillating behavior within the optical
rotation curves of some galaxies, for example NGC2998 \cite{rub3};
cf.~`ripples' in the light curve profile \cite{kent}. The explanation is that
across a spiral arm, positive velocity gradients are observed; the velocity
decrease from the outer edge of one arm to the inner edge of the next
arm is in some galaxies faster than Keplerian what is taken as compelling
evidence for noncircular velocities. In Fig.~\ref{fig5a}, we fit
the rotation curve data from Rubin et al.~\cite{rub3}. We have taken into
account four components: A truncated bulge, a thin disk of stars,
an HI component, and a dominating dark halo. What one can recognize is that
the data are not fit too well but that the maxima and minima of the fit
and the data are at the same place. The dark halo has to be the main part
in order to find a good fit. (The partly non-smooth behavior of the
curves results from numerical problems because of the scaling of our solution
by the scalar field frequency $\omega $.) Further investigations will
show whether cylindrically or axisymmetric solutions can improve the
fit or rule out our model. NGC2998 belongs to a sample of further 22
spiral galaxies for which the MOND model was applied \cite{sand}.
Oscillations (`wiggles') have been found also
in H$\alpha $ rotation curves \cite{marc}.

\section {\bf Scaling of the solutions and the mass distribution}

As we have seen from the redefined quantity $x= \omega r$, the frequency
$\omega $ scales the physical dimension of the solution.
Additionally, the mass is scaled with $1/\omega $ and the
energy density with $\omega^2$.
Table \ref{table1} shows for different values of $\omega $ the mass
and the density at different radii. 

There exist two ways how one can interpret the values of Table \ref{table1}.
The last row ($\omega =10^{-21}$/cm) shows a halo having a central
density of 10$^{-21}$g/cm$^3$; at $x=20$ which means $r=20$kLy the density
decreases to a value of 10$^{-23}$g/cm$^3$ and within a sphere of this
radius the halo has a mass of 10$^{43}$g$=10^{10} M_\odot $.

\begin{table}[b]
\caption[]{Order-of-magnitude estimation for $A=10^{-3}$.
We put the frequency $1/\omega =$
1cm, 10$^{15}$cm (about the radius of our solar system), 10$^{18}$cm = 1Ly,
and 10$^{21}$cm = 1kLy.
Further: $\omega $ in units of eV; density $\rho $ at the
center and at a distance of $x=20$; the mass within a sphere of radius
$x=20$.}
\begin{tabular}{dddddd}
 $\omega $ [1/cm] & $\omega $ [eV] & $\rho (0)$ [g/cm$^3$]
& $\rho (20)$ [g/cm$^3$] & $M (20)$ [g] \\
\tableline
 1          & 10$^{-5}$  & 10$^{21}$  & 10$^{19}$  & 10$^{24}$ \\
 10$^{-15}$ & 10$^{-20}$ & 10$^{-9}$  & 10$^{-11}$ & 10$^{37}$ \\
 10$^{-18}$ & 10$^{-23}$ & 10$^{-15}$ & 10$^{-17}$ & 10$^{40}$ \\
 10$^{-21}$ & 10$^{-26}$ & 10$^{-21}$ & 10$^{-23}$ & 10$^{43}$ \\
\end{tabular}
\label{table1}
\end{table}

But one can use Table \ref{table1} also in the vertical direction.
A solution with $\omega =1$/cm (first row) has within a sphere of
radius 20cm a mass of 10$^{24}$g, within a sphere of $20\times 10^{15}$cm
(20 times the diameter of our solar system; second row) a mass of
10$^{37}$g, within a sphere of 20Ly (third row) a mass of
10$^{40}$g, and so on. The same procedure is valid for the density columns.
The reason why one can do this is that one has a $\omega $ independence
of the density and the mass at high radii, but a $\omega $ dependence
for small radii. For the mass (\ref{mass}), we find
\be
M(r) = \frac{4\pi }{\kappa } A^2 \left \{ r +
 \frac{\cos (2 \omega r) - 1}{2 \omega^2 r} \right \} \; , \label{mass2}
\ee
which goes over for small $r$ values into
\be
M(r) \propto \frac{4\pi }{\kappa } A^2 \left \{ \frac{1}{3} \omega^2 r^3
- \frac{2}{45} \omega^4 r^5  \right \} \; .
\ee
The same is valid for the density $\rho $. This result shows that
the parameter $A$ determines the behavior of the halo at high distances
from the center (namely the limiting orbital velocity), but near the center
also the frequency of the scalar field influences the
behavior of the mass and the density. Rewriting Eq.~(\ref{mass2})
into
\be
2 \omega^2 r \left [\frac{\kappa M(r)}{4\pi } - A^2 r \right ] =
 \cos (2 \omega r) - 1
\ee
we immediately see that the left hand side has to be non-positive from which
we can find the inequality
\be
\frac{\kappa }{4\pi } \frac{M(R)}{R} \le A^2 \; ,
\ee
at an arbitrary radius $R$.
This inequality states that at every radius $R$, the relation between the
Schwarzschild radius and the radius of the object is smaller than the
square of the central value of the scalar field which is 10$^{-6}$
for a galaxy with $v_{max}=300$km/s.

Let us, for example, take the value $\omega =10^{-21}$/cm of the last row
of Table \ref{table1}. To calculate the mass and the density for smaller
radii, we have to take the formulae (\ref{rho}) and (\ref{mass}).
The result can be seen in
Table \ref{table2}. We recognize in comparison with the values of Table
\ref{table1} substantial smaller values.

\begin{table}[b]
\caption[]{Order-of-magnitude estimation for $\omega =10^{-21}$/cm.
In comparison with Table \ref{table1}, we find smaller densities and
masses at smaller distances.}
\begin{tabular}{lddddd}
 distance to center & $\rho $ [g/cm$^3$] & $M$ [g] \\
\tableline
 20 cm           & 10$^{-21}$ & 10$^{-17}$ \\
 20 10$^{-3}$ Ly & 10$^{-21}$ & 10$^{28}$  \\
 20 Ly           & 10$^{-21}$ & 10$^{37}$  \\
 20 kLy          & 10$^{-23}$ & 10$^{43}$  \\
\end{tabular}
\label{table2}
\end{table}

The order of magnitude values of Table \ref{table2} for the mass and
the density show also a possibility of a hierarchy of objects.
Recently \cite{ikebe}, it has been detected that the dark matter
distribution in the Fornax cluster of galaxies is a mixture of two
distinct components on different scales
(not necessary distinct matter components).
One part is associated with the cluster and the second smaller one with
the galaxy NGC1399.
This observation was done by looking into the X-ray-emitting diffuse
plasma of temperatures 10$^7$-10$^8$K which is widely found in elliptical
galaxies and clusters of galaxies. This gas is likely to be in hydrostatic
equilibrium, where the thermal pressure is balanced by the gravity
of the cluster. In this way, the X-ray emission traces the distributions
of the dark matter in regions of higher density where more gas is trapped.
This observation is the first evidence for a
hierarchical nature of the dark matter distribution. Future investigations
will show whether two of our dark matter halos can be combined:
one for the cluster, the other for the single galaxies.

\section {\bf Energy density, tangential pressure, and regularity}

\begin{figure}[ht]
\centering 
\leavevmode\epsfysize=7cm \epsfbox{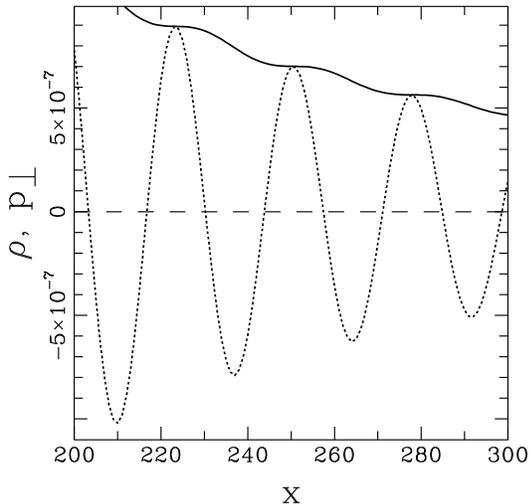}\\ 
\caption[]
{The energy density $\rho$ (solid curve) and the tangential pressure
$p_\bot $ (dotted curve) [both in units of $\omega^2/\kappa $] for the
interval of $x=200$ up to $x=300$. One recognizes very clearly the
terraced decrease of the energy density. The matter behaves like stiff
matter at the saddle points of $\rho $.
[Initial values: $\sigma (0)=1$ and $\nu(0)=1$]}
\label{fig6}
\end{figure}

An investigation of the energy density
$\rho $ (i.e.~the radial pressure $p_r$)
shows a terraced decrease (Fig.~\ref{fig6}).
Furthermore, we find the relation
$\mid \rho \mid$ $\ge $ $\mid p_\bot \mid $, where the tangential
pressure $p_\bot $ oscillates sinusoidally around zero.
From the equation of state, one recognizes that
at each extremum of the scalar field $\sigma $ the difference between radial
and tangential pressure vanishes. Hence, at constant spheres,
the equation of state $\rho = p$ of a {\em stiff} fluid arises.

We check numerically the quadratic curvature invariant
(Kretschmann scalar)
$
R_{inv}^2 = R^{\lambda \sigma \mu \nu } R_{\lambda \sigma \mu \nu }
$
revealing its global regularity for our solutions \cite{schunck}.
The same was found for the two invariants
$R_{\mu \nu } R^{\mu \nu }$ of the Ricci tensor and $R^2$ of the
curvature scalar. The invariants of the irreducible decomposition of
Riemann's curvature tensor show also no singularity \cite{schunck}.
Hence our solutions possess no physically relevant singularity.

\section {\bf Construction of a surface}\label{cons}

A massive complex scalar field with mass $m$ can be used for constructing
a surface of our halo. This massive scalar field has an exponentially
decrease and produces a finite mass. We have two different coordinates
$x_1=m r_1$ for the surface with the massive complex scalar field and
$x_2=\omega r_2$ in the interior with the massless complex scalar field.
At the connection we require $x_1=x_2$, hence $\omega /m = r_1/r_2$.
We need an equilibrium of the radial pressures and continuity of the metric
potentials. Because of the different scalings in the two regions, we
find for the radial
pressures $\omega (2)/m = \sqrt {p_r(1)/p_r(2)}$ [argument $(1)$, e.g.~in
$\omega (1)$, describes the model of massive complex scalar field and
argument $(2)$ that of the massless one], i.e., if we have the
equality of the metric potentials, the relation between the pressures
gives us the relation between the frequency of
the massless scalar field and the mass of the massive scalar field.
It was possible to find numerically such solutions, for example:
Initial values for the interior region:
$\sigma (0)=0.2333, \nu (0)=-0.4055$; for the surface region:
$\sigma (0)=0.1, \nu (0)=-0.14035418$; equality for the metric
potentials was looked for at $x=3$ and we find $\omega (2)/m=0.3203$.
Within the surface, the metric potentials of the massive scalar field
go over to the Schwarzschild metric. Hence, we have an object with a finite
mass. In the next Section, we shall show another possibility to
stop the halo.

\begin{table}[t]
\caption[]{This table shows the radius $R$ at which our dark matter halo
reaches the density $\rho $. $v$ is the limiting orbital velocity.}
\begin{tabular}{ddddd}
 $\sigma (0)$ & $v$ [km/s] & $\rho $ [g/cm$^3$] & $R$ [Mpc] \\
\tableline
$10^{-3}$ & 300 & 10$^{-29}$g/cm$^3$ & 2.3 \\
          &     & 10$^{-34}$g/cm$^3$ & 830 \\ \hline
$4\times 10^{-4}$ & 120 & 10$^{-29}$g/cm$^3$ & 0.9 \\
                  &     & 10$^{-34}$g/cm$^3$ & 330 \\
\end{tabular}
\label{table3}
\end{table}

\section {\bf Cover for the halo}\label{cover}

The energy density $\rho $ of the Newtonian solution decreases in
leading order as $A^2/x^2$. Therefore, we can come to the conclusion
that the energy density will eventually attain an hydrostatic equilibrium
with another matter form at the same pressure value, e.g.~the cosmic
microwave background radiation (CMBR), which is about 10$^{-34}$g/cm$^3$
(seas of neutrinos and supersymmetric particles
would increase this density).
Alternatively, a cosmological constant $\Lambda $ of the order of magnitude
10$^{-29}$g/cm$^3$ would produce much smaller halos; for estimates
of $\Lambda $ see \cite{car}. From Table \ref{table3}, we derive a radius
of about 10$^6$pc for $\Lambda $ and about 10$^8$pc for the CMBR value
(cf.~the discussion in \cite{rubin}). Recent observations show halo
radii of about 400kpc \cite{zar2}. Outside the halo, within the
spacetime with a constant density, we have a Schwarzschild-de-Sitter
metric.

\section {\bf Gravitational Redshift and rotation velocity}\label{gene}

By increasing the initial value of the scalar field $\sigma (0)$,
the Newtonian
description breaks down, i.e.~the formulae from Section \ref{newt}
are no longer valid. The metric potentials $\nu $ and
$\lambda $ are not constant and the full nonlinearity of Einstein's
theory has to be taken into account.

The mass and the density function change their behavior smoothly.
They still increase together with $\sigma (0)^2$ but one recognizes
an additional decrease of the $x$ dependent terms;
Table \ref{table4} shows the deviation from Newtonian behavior for the
mass function with increasing $\sigma (0)$. This means that one can
still use the order of magnitudes of Table \ref{table1}
(which was for $\sigma (0)=10^{-3}$) but one has to multiply the values
of the mass and the density by 10$^4$ for $\sigma (0)=0.1$ and
by 10$^6$ for $\sigma (0)=1$, for example. The deviation of the formula
is so eminent that one has to use numerical calculations.

\begin{table}[htb]
\caption[]{Deviation for $M/\sigma^2(0)$ from the Newtonian formula
(\ref{mass}) with increasing $\sigma (0)$; the mass still increases with
$\sigma (0)$. The values for $x=0.1$ have to be multiplied by
$10^{-3}$.}
\begin{tabular}{d|c|ccccc}
 $x$ & Newtonian & $\sigma (0)=0.1$ & 0.5 & 1.0 & 1.5 \\
\tableline
 0.1 & 0.33288 & 0.33287 & 0.33255 & 0.33156 & 0.32991 \\
 1.0 & 0.291 & 0.290 & 0.267 & 0.212 & 0.154 \\
 2.0 & 1.586 & 1.571 & 1.261 & 0.747 & 0.422 \\
 3.0 & 2.993 & 2.955 & 2.235 & 1.193 & 0.633 \\
 4.0 & 3.856 & 3.798 & 2.799 & 1.513 & 0.811 \\
 5.0 & 4.816 & 4.715 & 3.166 & 1.739 & 0.972 \\
\end{tabular}
\label{table4}
\end{table}

The strong curvature connected with these general relativistic (GR) solutions
produces also a gravitational redshift $z$ which, for a static
spherically symmetric mass distribution, is given by
\be
z = e^{[\nu (A_1)-\nu (A_0)]/2} - 1 \; ,
\ee
where $A_0$ corresponds to the place of the emitter and $A_1$ to that
of the receiver, both in rest. Table \ref{table5} shows results
for different values of $\sigma (0)$ up to $x=500$. The redshift increases
with growing $\sigma (0)$ and with growing radius.
Redshift values reached by a(n emitting) gas cloud at the center of the
halo can achieve very high values; we have not found numerically any limits.
The $z$ values for different $x$ show the gravitational redshift
which appears if our GR halo ends at $x$. In Table \ref{table6}, we
assume that the gas clouds are
in rest within the gravitational halo potential (more likely they circulate
around the center so that an additional Doppler redshift has to be taken
into account). Then, the redshifts can be interpreted as
emission or absorption lines of the corresponding gas clouds.
An interpretation of both Tables could be (for $\sigma (0)=1.0$):
Some radiation comes from the center and is gravitationally redshifted by
$z=4.76$, some part of the radiation is absorbed from a gas cloud at a
distance $x=1$ ($z=3.398$), another part at $x=100$ ($z=0.087$), and finally
the halo ends at $x=500$ so that gravitational redshift does not take
any further effect.
Near the center ($x=1$), we recognize a steep decrease
of redshift values (cf.~Table \ref{table5} and \ref{table6}).
The scale of such a halo is again defined by the frequency $\omega $.
If one assumes $\omega =10^{-18}$/cm, then the halo has a diameter of
$r=500$Ly; hence, not very far away from the source,
about $r=400$Ly, one has already small gravitational redshift values.

\begin{table}[t]
\caption[]{The redshift function $z(x)=\sqrt{\exp [\nu (x)-\nu (0)]} -1$
for solutions with different initial values
of the scalar field ($\nu (0)=0$ in each case).}
\begin{tabular}{d|ccccccc}
$z$     & $\sigma (0)=$ 0.25 & 0.5 & 0.75 &
 1. & 1.25 & 1.5 \\
\tableline
$z$(10)  & 0.16 & 0.66 & 1.58 & 3.16 & 5.88 & 9.52  \\
$z$(50)  & 0.25 & 0.97 & 2.13 & 4.04 & 7.83 & 16.17 \\
$z$(100) & 0.29 & 1.08 & 2.31 & 4.29 & 8.29 & 17.57 \\
$z$(200) & 0.32 & 1.18 & 2.47 & 4.50 & 8.66 & 18.65 \\
$z$(300) & 0.35 & 1.24 & 2.56 & 4.62 & 8.86 & 19.17 \\
$z$(400) & 0.36 & 1.28 & 2.61 & 4.70 & 8.98 & 19.52 \\
$z$(500) & 0.37 & 1.31 & 2.66 & 4.76 & 9.08 & 19.77 \\
\end{tabular}
\label{table5}
\end{table}

\begin{table}
\caption[]{Redshift values for absorption lines
$z(x)=\sqrt{\exp [\nu (500)-\nu (x)]} -1$
for different initial values
of the scalar field ($\nu (0)=0$ in each case).
We assume that the halo ends at $x=500$.}
\begin{tabular}{d|ccccc}
$z$     & $\sigma (0)=$ 0.25 & 0.5 & 0.75 & 1.  \\
\tableline
$z$(1)   & 0.350 & 1.147 & 2.124 & 3.398  \\
$z$(5)   & 0.229 & 0.536 & 0.630 & 0.669  \\
$z$(10)  & 0.183 & 0.386 & 0.417 & 0.383  \\
$z$(50)  & 0.096 & 0.170 & 0.168 & 0.141  \\
$z$(100) & 0.064 & 0.108 & 0.105 & 0.087  \\
$z$(200) & 0.034 & 0.056 & 0.054 & 0.045  \\
$z$(300) & 0.019 & 0.030 & 0.029 & 0.024  \\
$z$(400) & 0.008 & 0.012 & 0.012 & 0.010  \\
$z$(500) & 0.    & 0.    & 0.   & 0.    \\
\end{tabular}
\label{table6}
\end{table}

\begin{figure}
\centering 
\leavevmode\epsfysize=8.5cm \epsfbox{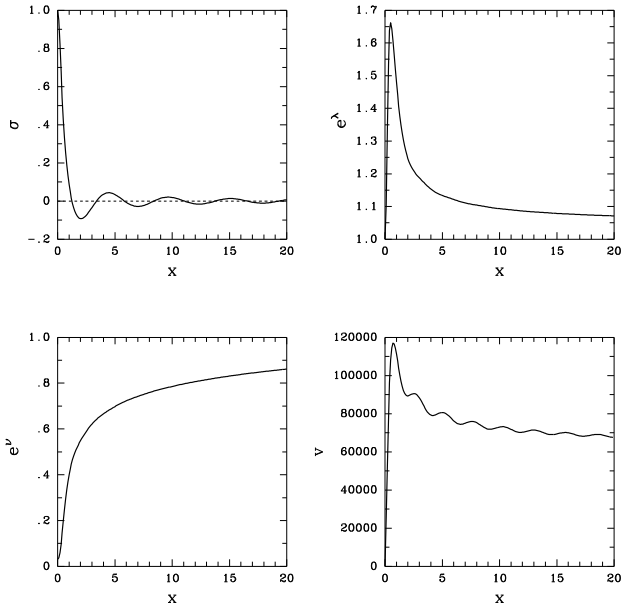}\\ 
\caption[]
{The scalar field $\sigma $, the metric potentials $e^{\nu }$ and
$e^{\lambda }$, and the rotation velocity $v$ (in units of [km/s])
for high relativistic initial values of $\sigma (0)=1$ and $\nu (0)=-3.5$.
Near the center, particles in circular orbits have velocities of more
than one third of the velocity of light.}
\label{fig7}
\end{figure}

How fast does a mass rotate in such a spacetime? Figure \ref{fig7} shows
a numerically calculated solution. We see that velocities of more than
10$^5$km/s are reached. This means that 6\%\ of the rest mass energy is
stored in kinetic energy. If we assume that per year a mass of 1$M_\odot $
transfers its kinetic energy into radiation then we find a luminosity
of 10$^{44}$erg/s.

\section {\bf Discussion}

We have shown that we were able to fit the universal rotation curves of
Persic and Salucci and of a selection of dwarf and spiral galaxies which are
especially suitable to find out characteristics of isolated dark matter halos.
In one case, we could match oscillations of rotation curve data and
our model.

We have to use Newtonian solutions and to fix two parameters.
There is the central amplitude of the scalar field, the parameter $A$,
which determines the limiting orbital velocity.
The second parameter, the frequency of the scalar field, $\omega $,
varies the values of the mass and the density near the center of the
galaxies.

Because our solution consists of a condensed `star'-like object
(with increasing mass) also the energy contribution of the radial
pressure $p_r$ plays a role for the rotation velocity. We found that
this part contributes the same order of magnitude as
the Newtonian part coming from the `normal' energy density $\rho $.
Therefore, general relativity is just the correct theory for the dark
matter part provided a compact object with internal pressure
is present.

Another issue is the physical basis of the Tully-Fisher relation.
It is obvious that in a deeper gravitational potential, i.e.~more dark
matter, also more luminous matter can find place.
Our best fits for the rotation curves are just found if we
add for low luminosity galaxies an almost dominating scalar field halo
and vice versa. The contribution of our dark matter form is determined by
the parameter $A$ the values of which lie in the order of magnitudes
10$^{-4}$ and 10$^{-3}$.
In a future paper, we will investigated why such values are preferably found.

In a recent paper \cite{rhe}, the reason for
the Tully-Fisher relation is divided into two parts: (a) a mass-to-light
ratio of the luminous matter and (b) a relation between the luminous mass
and the rotation velocity. It is shown that there is a smaller scatter
in part (b) than (a), because (a) depends on the present star formation
rate. The conclusion is that part (b) in combination with a well-behaved
relation between luminous and dark matter (producing flat and smooth
rotation curves, i.e.~the `conspiracy') is the physical basis of the
Tully-Fisher relation.
The conspiracy maintains a luminous-mass-rotation-velocity
relation with a slope of 4. The issue is whether one can reveal a reason
from our model. We will investigate this point in a future work.

Providing this halo model is realized in nature and one finds a radius
of the dark matter halos of galaxies, then this radius could
be a hint for a cosmological constant and its order of magnitude.

During the completion of this paper investigations with massive scalar
particles were carried out. In \cite{deh1,deh2}, an isothermal ideal
Bose gas which is degenerated in the center of galaxies was used to
fit rotation curves of 36 spiral galaxies. The authors found the best
fits with a boson particle mass of about 60eV.
In \cite{sin} excited Newtonian boson star solutions and in \cite{koh}
excited general relativistic boson stars were used as
dark halos. In the latter case, the contribution of the radial pressure
to the rotation velocity was wrongly neglected. It is well-known that
excited boson stars are unstable against small radial perturbations
(see e.g.~\cite{kusmartsev}); for a recent numerical investigation
of instability of excited boson stars, see \cite{bala1}. Under the
assumption that boson stars are transparent (as the solutions of this
paper), maximal gravitational redshift values of about 0.7 have been revealed
\cite{lid}.
In \cite{bala2}, it was shown numerically and analytically that the
Newtonian solutions of our model are stable.

\acknowledgments
We would like to thank Peter Baekler, John D.~Barrow, Heinz Dehnen,
Ralf Hecht, Friedrich W.~Hehl, Martin Hendry,
Andrew R.~Liddle, Eckehard W.~Mielke, Yuri N.~Obukhov, Jochen Peitz,
Domingos da C.~Rodrigues, Vera C.~Rubin, Paolo Salucci, Peter Thomas,
and Pedro T.P.~Viana for helpful discussions and
comments. The referees have improved the contents of the
paper with their recommendations.
Research support was provided by an European Union Marie Curie
TMR fellowship.



\end{document}